\documentclass{aastex}
\usepackage{emulateapj5}
\shorttitle{B\"ohm--Vitense Gaps in the Hyades}
\shortauthors{J.H.J.\ de Bruijne, R.\ Hoogerwerf, \& P.T.\ de Zeeuw}
\begin{document}
%
%
\title{Two B\"ohm--Vitense gaps in the main sequence of the Hyades}
\author{J.H.J.\ de Bruijne\altaffilmark{1}, R.\ Hoogerwerf, \& P.T.\ de Zeeuw}
\affil{Leiden Observatory}
\affil{PO Box 9513, 2300 RA Leiden, The Netherlands}
\email{jdbruijn@astro.estec.esa.nl}
\altaffiltext{1}{Present address: Astrophysics Division, ESA, ESTEC, PO Box 299, 2200 AG Noordwijk, The Netherlands}

%
%
\begin{abstract}
Hipparcos proper motions and trigonometric parallaxes allow the
derivation of secular parallaxes which fix the distances to individual
stars in the Hyades cluster to an accuracy of $\sim$2~percent. The
resulting color-absolute magnitude diagram for 92 high-fidelity single
members of the cluster displays a very narrow main sequence, with two
turn-offs and associated gaps. These occur at the locations where the
onset of surface convection affects the $B-V$ colors, as predicted by
B\"ohm--Vitense thirty years ago. The new distances provide stringent
constraints on the transformations of colors and absolute magnitudes
to effective temperatures and luminosities, and on models of stellar
interiors.
\end{abstract}
%
%
%
%
\section{Introduction}
The Hyades open cluster is one of the key calibrators of the absolute
magnitude-spectral type relation and the mass-luminosity relation.
The small distance to the cluster ($\sim$45~pc) has a number of
advantages for studies of its Hertzsprung--Russell diagram: (i) the
foreground interstellar reddening and extinction is negligible
($E(B-V) = 0.003 \pm 0.002$~mag \citep[e.g.,][]{tay1980}), (ii) the
large proper motion ($\mu \sim 111$~mas~yr$^{-1}$) and peculiar space
motion ($\sim$35~km~s$^{-1}$) facilitate proper motion- and radial
velocity-based membership determinations, and (iii) the stellar
content can be probed to low masses relatively easily.  Among the
$\sim$400 known members are white dwarfs, red giants, mid-A stars in
the turn-off region, and main-sequence stars down to at least
$\sim$$0.10~M_\odot$ M dwarfs \citep*{bhj1994}. At the same time, the
proximity of the Hyades complicates astrophysical interpretation: the
tidal radius of $\sim$10~pc corresponds to a significant depth along
the line of sight. As a result, the precise definition and location of
the main sequence and turn-off region in the Hertzsprung--Russell
diagram, and the accuracy of the determination of, e.g., the helium
content and age of the cluster, have always been limited by the
measurement error of the distances to individual stars. Even Hipparcos
parallax uncertainties translate into absolute magnitude errors of
$\ga$0.10~mag at the mean distance of the cluster, whereas $V$-band
photometric errors only account for $\la$0.01~mag uncertainty for most
members.

\begin{figure*}[t]
\begin{center}
\includegraphics[angle=0.0, width=18.0truecm, 
                 clip=true, keepaspectratio=true]
                 {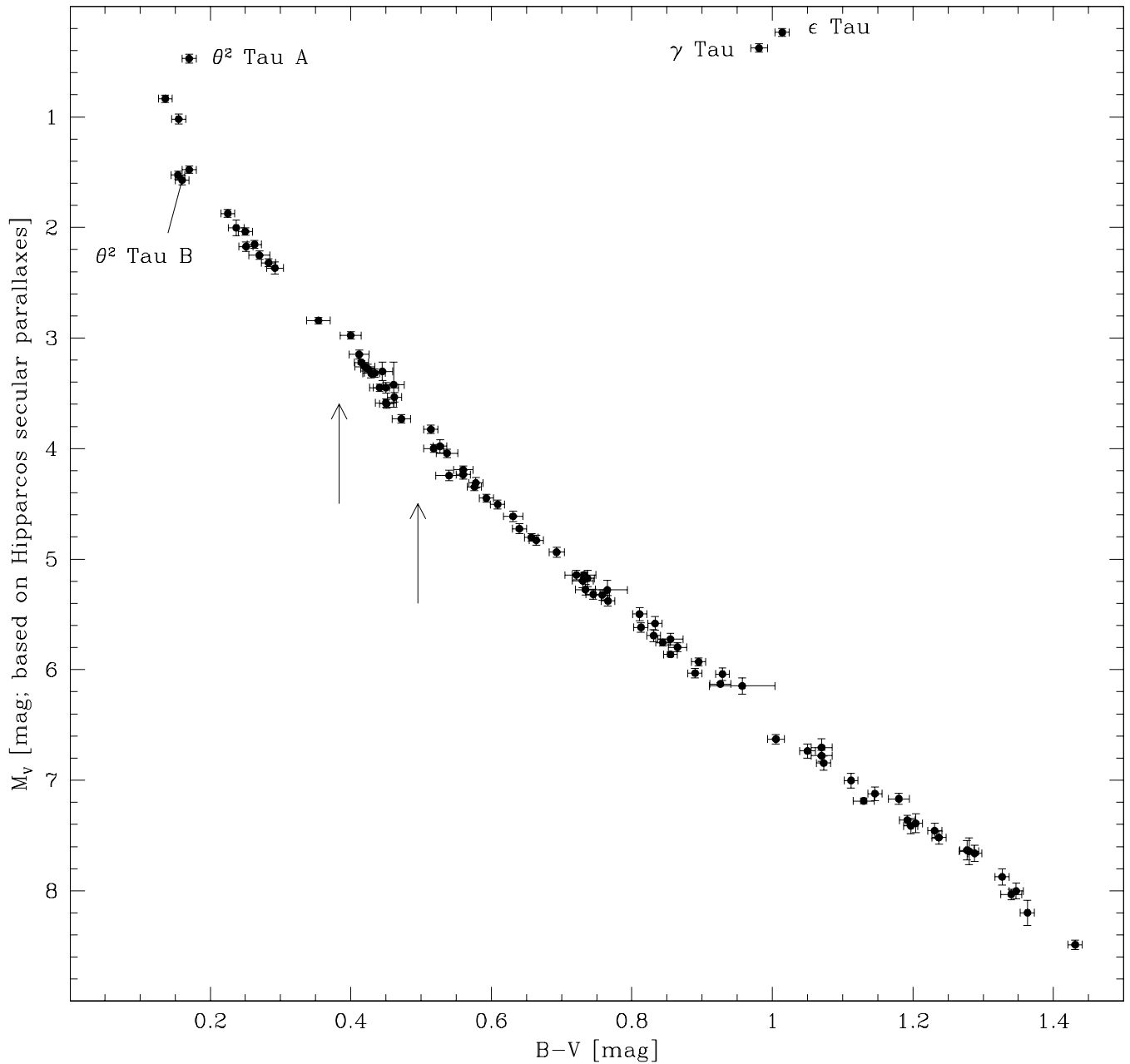}
\end{center}
  \caption[]{Color-absolute magnitude diagram for 92 high-fidelity
members of the Hyades cluster. This sample excludes all members beyond
40~pc from the cluster center, double and multiple stars, and stars
with suspect secular parallaxes (see BHZ for details). The absolute
magnitudes were computed using the observed $V$-band magnitudes and
secular parallaxes derived by \citet{bhz2000}. The $B-V$ colors were
taken from the Hipparcos Catalog. The arrows indicate the two
B\"ohm--Vitense turn-offs and associated gaps, which are most likely
caused by sudden changes in the properties of convective
atmospheres. The gaps between $B-V=0.15$ and $0.20$, between
$B-V=0.30$ and $0.35$, and the gap around $B-V=0.95$ mag are caused by
the suppression of double, multiple, and peculiar stars from our
sample (cf.\ figure 21 in \citet{per1998}; see \S 2); the region
between $B-V=0.30$ and $B-V=0.35$ mag, e.g., is occupied by Am-type
stars, which have a high incidence of duplicity
\citep[e.g.,][]{jj1987}.}
\label{fig:camd}
\end{figure*}

It was realized long ago \citep[e.g.,][]{bos1908} that the small
internal velocity dispersion of the Hyades makes it possible to
improve the trigonometric parallaxes of individual member stars by
kinematic modelling of their proper motions. Various recent studies
derived such secular parallaxes for Hyades members from Hipparcos
astrometry \citep*{per1998,bru1999,dlm1999,ng1999}, but none
investigated the resulting Hertzsprung--Russell diagram in any
detail. Yet, as the relative proper motion accuracy is effectively
three times larger than the relative trigonometric parallax accuracy,
the secular parallaxes are three times more precise than the
trigonometric parallaxes, and provide a unique opportunity to obtain a
well-defined and absolutely calibrated Hertzsprung--Russell diagram
for an open cluster. For this reason we have redetermined the secular
parallaxes for the Hyades, using the comprehensive procedure described
by \citet{bru1999} \citep*[cf., e.g.,][]{lmd2000}. We describe the
full derivation of the secular parallaxes elsewhere
\citep*[BHZ]{bhz2000}. This includes a new analysis of the space
motion and internal velocity dispersion of the Hyades, a validation of
the secular parallaxes, with a detailed investigation of the effect of
velocity structure in the cluster and of the presence of
small-angular-scale correlations in the Hipparcos data, and an
extensive analysis of the construction of the physical
Hertzsprung--Russell diagram $(\log T_{\rm eff}, \log L)$.  In the
course of this work, we constructed a color-absolute magnitude diagram
for a sample of high-fidelity single members of the Hyades, and
discovered that the improved accuracy reveals two turn-offs and
associated gaps in the main sequence at the locations predicted by
B\"ohm--Vitense, in the seventies. These are the topic of this Letter.

\begin{figure*}[t]
\begin{center}
\includegraphics[angle=-90.0, width=8.0truecm, 
                 clip=true, keepaspectratio=true]
                 {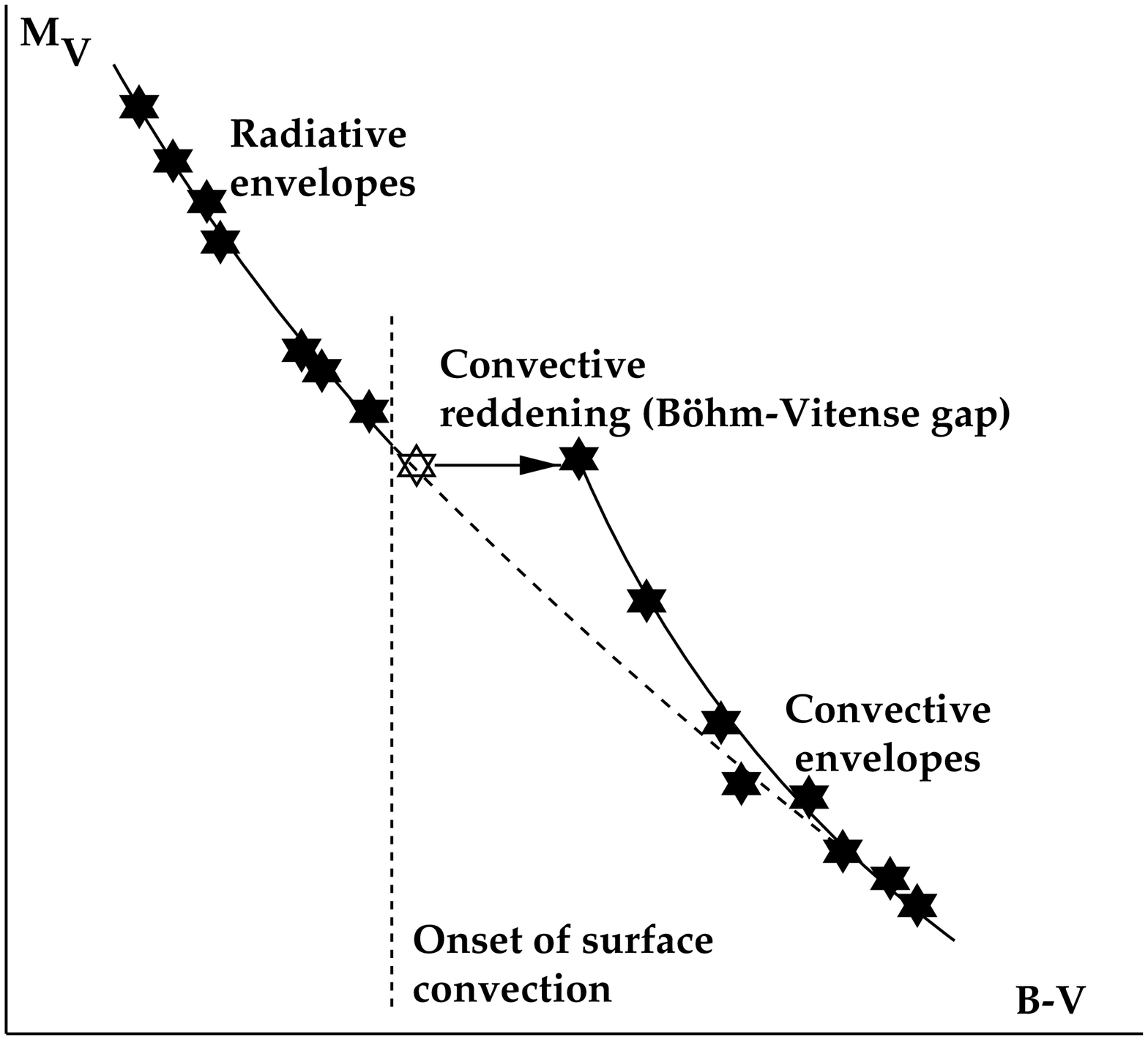}
\end{center}
  \caption[]{Schematic color-absolute magnitude diagram of an open
cluster with a B\"ohm-Vitense gap. Stars left of the dashed vertical
line, at $B-V\sim 0.3$ mag, have radiative envelopes. Surface
convection sets in at the line, and all stars to the right of it have
convective envelopes. While the onset of convection does not alter
luminosity or effective temperature, it causes redder colours because
of lower temperatures in deeper layers (see \S 3). The resulting
color-absolute magnitude diagram therefore shows a turn-off and gap.}
\label{fig:bvgap}
\end{figure*}

\section{The $(B-V,M_V)$ diagram}\label{sec:camd}

The Hipparcos Catalog (ESA 1997) contains about 200 members of the
Hyades brighter than $V\sim12.4$ (\citealt{per1998}; BHZ).  We
selected a high-fidelity subset of single members, defined as
follows. We removed suspect kinematic members and stars which have
known spectral peculiarities and hence deviant positions in the
color-magnitude diagram.  We excluded the stars beyond 40~pc from the
cluster center (for which membership is not particularly certain) and
all (possible) close double and multiple systems (98 spectroscopic
binaries, Hipparcos DMSA--`G$\,$O$\,$V$\,$X$\,$S' stars, and stars
with suspect secular parallaxes). We also rejected 11 stars which are
variable (Hipparcos field H52 is one of `D$\,$M$\,$P$\,$R$\,$U') or
have large photometric errors ($\sigma_{B-V} > 0.05$~mag), as well as
the suspect stars HIP 20901, 21670, and 20614 (see BHZ for further
details).

The final sample contains 90 single members which follow the main
sequence from $B-V \sim 1.43$~mag (late-K/early-M dwarfs; $M \ga
0.5~M_\odot$) to $B-V \sim 0.10$~mag (A7IV stars; $M \sim
2.4~M_\odot$), except for $\epsilon$ and $\gamma$ Tau, which are
evolved red giants. We added the two components of the `resolved
spectroscopic binary' $\theta^2$~Tau as single stars. These are
located in the turn-off region of the cluster, and bring the total
number~of objects to~92. We took the accurate values for $B-V$ and $V$
listed in the Hipparcos Catalog, and used the secular parallaxes to
compute $M_V$.

Figure~\ref{fig:camd} shows the resulting color versus secular
parallax-based absolute magnitude diagram for these 92 stars
\citep[cf.][]{mad1999}. It shows the giant clump, the turn-off region,
and a well-defined and very narrow main sequence. The latter contains
a few conspicuous but artificial gaps, e.g., around $B-V=0.95$ and
between $B-V=0.30$ and $B-V=0.35$ mag, caused by the suppression of
(double and multiple) stars from our sample \citep[cf.\ figure 21
of][]{per1998}. The small gap between $B-V=0.75$ and $0.80$ mag is
also present in the original sample \citep[see figure 21
of][]{per1998} and is therefore probably real \citep[cf.\ figure 2
of][]{mer1981}. Somewhat further to the blue, there are two turn-offs
and associated gaps, near $B-V \sim 0.38$ and $\sim$$0.48$~mag. These
features are not clearly discernible in the lower-quality
trigonometric parallax-based version of the same diagram \citep[figure
21 of][]{per1998}. Although their reality in the `cleaned' secular
parallax colour-magnitude diagram is hard to establish beyond all
doubt, the simultaneous existence of both a turn-off and an associated
gap at a location which coincides with predictions made by stellar
structure models (see \S 3) strongly argues in favour of them being
real \citep*[cf.][and references therein]{k&f1991}.

\section{B\"ohm--Vitense gaps}

\citet{boh1970,boh1981,boh1982} realized that convective atmospheres
have relatively low temperatures in the deeper layers which contribute
to the surface flux in the spectral regions of the $U$ and $B$ filters
\citep{nel1980}. As a result, convective atmospheres appear to be
cooler, and therefore have $B-V$ colors that are reddened by amounts
of $\sim$0.07--0.12~mag when compared to radiative atmospheres of the
same $T_{\rm eff}$. As the reddening of the atmosphere is not
accompanied by a significant change in luminosity, the onset of
surface convection can therefore cause a turn-off and associated gap
in the color-absolute magnitude diagram, starting around $B-V =
0.25$--$0.35$~mag (Figure~\ref{fig:bvgap}). The precise location of
this B\"ohm--Vitense gap depends on details of the stellar structure,
including metallicity. We identify the turn-off and gap at
$B-V\sim0.38$ in Figure~\ref{fig:camd} as the B\"ohm--Vitense gap in
the main sequence of the Hyades.

Observational evidence for a B\"ohm--Vitense gap in the main sequence
is sparse \citep[e.g.,][]{bvc1974,jas1984,rc2000}, and its reality has
been disputed \citep[e.g.,][]{mp1988,sl1997,ny1998}. Previous claims
for its existence were based on the presence of gaps in either
color-color diagrams or in the cumulative distribution of cluster
members in some photometric index \citep*[e.g.,][]{adm1969}, instead of
on the presence of `gaps' or `turn-offs' in color-absolute magnitude
diagrams. The high-accuracy color-absolute magnitude diagram presented
in Figure~\ref{fig:camd} provides the first direct evidence for the
existence of the B\"ohm--Vitense turn-off and gap.

The presence of a gap near $B-V \sim 0.48$~mag ($T_{\rm eff} \sim
6400$~K) in the main sequence of the Hyades was already noted by
\citet{boh1995a,boh1995b}, who attributed it to `{\it a
sudden increase in convection zone depths}'. Figure~\ref{fig:camd} reveals
the associated turn-off. The position of this second B\"ohm--Vitense
gap coincides with both the so-called Lithium gap, which is generally
thought to be related to the rapid growth of the depth of the surface
convection zone with effective temperature decreasing from
$\sim$7000~K to $\sim$6400~K
\citep[e.g.,][]{bt1986,mic1986,sfr1994,bal1995}, and with the onset of
dynamo-induced magnetic chromospheric activity
\citep*[e.g.,][]{wbs1986,grb1993}.

\section{Conclusions and prospects}\label{conclusions}

Narrow main sequences in color-magnitude diagrams are readily
observable for distant clusters, but the absolute calibration of the
Hertzsprung--Russell diagram of such groups is often uncertain due to
their poorly determined distances and the effects of interstellar
reddening and extinction. The latter problems are alleviated
significantly for nearby clusters, but at the price of introducing a
considerable spread in the location of individual members in the
Hertzsprung--Russell diagram as a result of their resolved intrinsic
depths, and relatively poorly determined individual distances. We have
shown here that the Hyades are unique in that the secular parallaxes
derived from the Hipparcos astrometry for the members are sufficiently
accurate to calibrate the main sequence of this nearest open cluster
in absolute terms.

The color-absolute magnitude diagram for the Hyades shown in
Figure~\ref{fig:camd} reveals two turn-offs and associated gaps in the
main sequence. We identify these with the so-called B\"ohm--Vitense
gaps, which are most likely related to sudden changes in the
properties of surface convection zones in the atmospheres of stars
with $B-V \sim 0.30$ and $\sim$0.45~mag, which significantly affect
the emergent UV and blue-optical fluxes, and thus the $B-V$ color. We
show in BHZ that this substructure in the $(B-V, M_V)$ diagram
provides a strong constraint on stellar models, requiring an
improvement in the available transformations from colors and absolute
magnitudes to effective temperatures and luminosities.

The future astrometric satellites FAME and GAIA will improve the
accuracy of stellar astrometry into the micro-arcsecond regime. They
will make it possible to provide accurate membership and
high-precision absolutely-calibrated main sequences for all star
clusters and associations to distances of at least 2 kpc. This will
provide the ability to test stellar models over a range of
metallicities and ages, and may well reveal further substructure.

\acknowledgements 
It is a pleasure to thank Adriaan Blaauw, Jan Lub, and Michael
Perryman for helpful comments and suggestions.

\end{document}